\documentclass[aps,preprint,showpacs,superscriptaddress,groupedaddress]{revtex4}
\usepackage{bm}       
\usepackage{amssymb}
\usepackage{amsmath}     

\usepackage{hyperref} 
\usepackage{psfrag}
\usepackage[cp1250]{inputenc}

\usepackage{graphicx}

\DeclareMathAlphabet{\mathpzc}{OT1}{pzc}{m}{it}

\newcommand{\mh}{$ 125 \, \textrm{GeV} \ $}
\newcommand{\mt}{$ 173.35 \, \textrm{GeV} \ $}

\def\gev{\,{\rm GeV}}
\def\tev{\,{\rm TeV}}

\begin{document}

\title{\boldmath Higgs boson mass and high-luminosity LHC \\probes of supersymmetry with vectorlike top quark \\}
\author{Zygmunt Lalak} 
\affiliation{Institute of Theoretical Physics, Faculty of Physics, University of Warsaw ul. Pasteura 5, 02-093 Warsaw, Poland}
\author{Marek Lewicki}
\affiliation{Institute of Theoretical Physics, Faculty of Physics, University of Warsaw ul. Pasteura 5, 02-093 Warsaw, Poland}
\affiliation{Michigan Center for Theoretical Physics, University of Michigan, Ann Arbor MI 48109, USA} 
\author{James D. Wells} 
\affiliation{Michigan Center for Theoretical Physics, University of Michigan, Ann Arbor MI 48109, USA} 

\begin{abstract}
We consider an extension of the MSSM with an added vectorlike top partner. Our aim is to revisit to what extent such an extension can raise the Higgs boson mass through radiative corrections and help ameliorate the MSSM hierarchy problem, and to specify what experimental probes at the LHC will find or exclude this possibility during the high-luminosity phase. Direct detection, precision electroweak and precision Higgs analyses are all commissioned to this end. To achieve the \mh Higgs boson mass, we find that superpartner masses can be reduced by a factor of more than three in this scenario compared to the MSSM without the extra vectorlike top quark, and that during the high-luminosity phase of the LHC precision Higgs analysis is expected to become the most powerful experimental probe of the scenario.
\end{abstract}

\date{\today}

\maketitle

\tableofcontents

\section{Introduction}

Despite the lack of experimental confirmation from the first run of LHC, supersymmetry  is still a promising solution to the Standard Model (SM) hierarchy problem. In the minimal realization of supersymmetry, the Higgs boson mass at tree level is bounded by the $Z$ boson mass and needs to be lifted up by radiative corrections from superpartners. This calls for large superpartner masses that introduce a new hierarchy between the weak scale and the scale of supersymmetry. This is often called the little hierarchy problem of the MSSM. 
   
We will focus on an extension of MSSM with a vectorlike top quark partner. This is the simplest of vectorlike matter extensions \cite{Moroi:1991mg,Moroi:1992zk,Babu:2004xg,Babu:2008ge,Martin:2009bg,Graham:2009gy,Martin:2010dc,Endo:2011mc,Faroughy:2014oka,Ellis:2014dza}
that can effectively reduce the little hierarchy due to large new contributions it induces to the Higgs mass.
To illustrate this point we use the simplest possible supersymmetry spectrum with all soft terms at the scale $M_{SUSY}$. The only exceptions are the $A$ terms equal to $-M_{SUSY}$. Also, the Higgs boson soft masses and $B$ parameters are chosen to accommodate correct electroweak symmetry breaking.
To this very simple spectrum we add a vectorlike top multiplet, $t'$ and $\bar t'$, where $t'$ has the quantum numbers of the right-handed top quark $t^c_R$ and $\bar t'$ is its conjugate. The soft masses of the scalar components of $t'$ and $\bar t'$ are also equal to $M_{SUSY}$. We include these new superfields into the superpotential and calculate  the contribution to the Higgs boson mass. As shown previously in different contexts as well~\cite{RaiseHiggs}, the addition of vectorlike states that mix with the MSSM fermions can raise the Higgs boson mass, thereby enabling smaller superpartner masses to achieve $m_h=125\gev$ through these additional radiative corrections.

We then calculate possible experimental exclusions or detections coming from precision electroweak measurements, corrections to Higgs boson properties and direct detection of the new vectorlike state. Finally we compare the impact of all these bounds on our model and calculate the lowest possible $M_{SUSY}$ consistent with these bounds.   

One key result is that the most constraining of the three experimental analyses is usually the modification of Higgs boson properties, except when there is  large $\tan \beta$ and small mixing. In that case, the direct detection of exotic vectorlike states at the LHC can be more probing. We also will show that when including all the constraints, $M_{SUSY}$ can still be lowered $3$ to $5$ times compared to the MSSM and still yield $m_h=125\gev$. Thus even a very simple vectorlike quark extension can significantly ameliorate the little hierarchy problem of MSSM.
Its important to point out that since we do not consider a specific UV completion, our measure of the little hierarchy problem is simply the splitting between the electroweak scale and the SUSY scale, rather than a result of some specific fine-tuning measure.

\section{MSSM with vectorlike top partner}

The superpotential of the MSSM with an additional vectorlike top partner (omitting small Yukawa couplings of the first two families), reads
\begin{equation}
W=Y_t Q H_u \bar{t}+Y_{t'} Q H_u \bar{t}'+m t' \bar{t} +M t' \bar{t}'+Y_b Q H_d \bar{b}+Y_\tau L H_d \bar{\tau}+\mu H_u H_d .
\end{equation}
The above superpotential leads to the following mass matrix in the basis $\Psi=(Q,t',\bar{t}^\dagger,\bar{t'}^\dagger)$:
\begin{equation}\label{fermionmass}
{\mathbf{M}_t}=
\left(
\begin{array}{cc}
0 & \mathbf{m}_t \\
 \mathbf{m}_t^\dagger&0\\
\end{array}
\right),
 \quad
{\mathbf{m}_t}=
\left(
\begin{array}{cc}
Y_t v_2 & Y_{t'} v_2 \\
 m&M\\
\end{array}
\right),
\end{equation} 
where $v=\sqrt{v_1^2+v_2^2}\approx 246$, $\tan{\beta}=v_2/v_1$ and $v_2= v \sin{\beta}/\sqrt{2}$.

In order to obtain masses of the fermions we diagonalize the mass matrix by unitary $L$ and $R$ matrices:
\begin{equation}\label{LRmatrixdef}
L {\mathbf m}_t R^{\dagger} = {\rm diag}(m_{t_1}, m_{t_2}).
\end{equation}
We always set the first eigenvalue equal to the top quark mass, while the second is the mass of the new vectorlike quark.

The mass matrix of the scalars takes the following form:
\begin{equation}\label{scalarmass}
{\mathbf{M}^2_S}={\mathbf{M}^2_t}+
\left(
\begin{array}{cccc}
m^2_{Q_3}+D_{\frac{1}{2},\frac{2}{3}}&0&\frac{v_u}{\sqrt{2}} A_{t} - \frac{v_d}{\sqrt{2}} \mu Y_t&\frac{v_u}{\sqrt{2}} A_{t'} - \frac{v_d}{\sqrt{2}} \mu Y_{t'} \\
0& m^2_{\bar{t}'}+D_{0,\frac{2}{3}}&B_m&B_M\\
\frac{v_u}{\sqrt{2}} A_{t} - \frac{v_d}{\sqrt{2}} \mu Y_t&B_m&m^2_{U_3}+D_{-\frac{1}{2},-\frac{2}{3}}&0\\
\frac{v_u}{\sqrt{2}} A_{t'} - \frac{v_d}{\sqrt{2}} \mu  Y_{t'} &B_M&0& m^2_{t'}+D_{0,-\frac{2}{3}}\\
\end{array}
\right),
\end{equation} 
 in the basis $\Phi=(\tilde{t},\tilde{t}',\tilde{\bar{t}},\tilde{\bar{t}}')$, where $D_{T_3,q}=(T_3-q\sin \theta_W )\cos(2 \beta) M_Z^2$ is the electroweak $D$ term contribution, and $A$ and $B$ are soft breaking terms corresponding to the appropriate couplings in the superpotential.
Due to mixing with the vectorlike quark, the top Yukawa coupling can now be very different from its  MSSM value while still keeping the predicted top mass unchanged. There are always two values of the top Yukawa that predict the correct top mass, and we always chose the larger one. The smaller value is a modification of the fermiophobic Higgs coupling approach, and generally is more constrained by the data.

In what follows we consider two sets of new parameters. One set incorporates the small mixing example with $m=0$, and the other incorporates the large mixing case with $m=M_{SUSY}$. In both cases the superpotential vectorlike mass term $M$ is also equal to $M_{SUSY}$. New scalar soft masses are $m^2_{\bar{t}'}=m^2_{t'}=M^2_{SUSY}$ and all other mass parameters which were not present in the MSSM are set to $B_m=B_M=A_{t'}=0$. For simplicity we set the pseudoscalar mass $m_A$ and all MSSM soft breaking terms to $M_{SUSY}$ except $m_{H_1}$, $m_{H_2}$ and $B$ which we vary in order to achieve correct electroweak symmetry breaking for each value of $M_{SUSY}$. A-terms are all set to $-M_{SUSY}$.  As mentioned above $Y_{t}$ is always fixed by requiring that at the tree level $m_{t_1} =m_t^{MSSM}$ which corresponds to the physical top mass $m_{t}=\, $\mt when one-loop corrections are included. The only free parameters left are $M_{SUSY}$ and $\tan{\beta}$.

\subsection{Higgs mass correction}

We calculate the contribution to the mass of the light neutral Higgs boson using effective potential approximation in the decoupling regime~\cite{Martin:2009bg}. The contribution to the effective potential from tops and stops and the new vectorlike states reads 
\begin{equation}\label{deltaveff}
\Delta V =\frac{6}{64 \pi^2}\sum_{i=1}^4  \left[ F(m^2_{\tilde{t}_i})-2F(M^2_{t_i}) \right]
\end{equation}
where $F(x)=x^2\ln (x/Q^2)$ while $M^2_{t_i}$ and $m^2_{\tilde{t}_i}$ are eigenvalues of the fermion mass matrix \eqref{fermionmass} and scalar mass matrix \eqref{scalarmass} respectively. The correction to the light Higgs boson squared mass is equal to 
\begin{small} 
\begin{equation}\label{higgsmasscorrection}
\Delta m^2_h =\left[
	 \frac{\sin^2 \beta}{2}\left( \frac{\partial^2}{\partial v^2_u}-\frac{1}{ v_u}\frac{\partial}{\partial v_u} \right)
	+ \frac{\cos^2 \beta}{2}\left( \frac{\partial^2}{\partial v^2_d}-\frac{1}{ v_d}\frac{\partial}{\partial v_d} \right)
	+\sin \beta \cos \beta \left( \frac{\partial^2}{\partial v_d \partial v_u} \right)
 \right]\Delta V.
\end{equation}
\end{small}
Since the above correction already includes the top and stop contribution, we subtract the MSSM top and stop correction $\Delta {m_h}^{\textrm{MSSM}}$ which was already included (among other corrections \cite{Higgsmass}) in our MSSM value $m_h^{\textrm{MSSM}}$. We calculate the $\Delta {m_h}^{\textrm{MSSM}}$ correction using eigenvalues of the MSSM mass matrices in equation \eqref{deltaveff} and then using an equation similar to \eqref{higgsmasscorrection}, with only MSSM masses.
Our final computation of the corrected Higgs mass reads
\begin{equation}\label{higgsmass}
m_h^2 =(m_h^{MSSM})^2+\Delta m_h^2-(\Delta m_h^{MSSM})^2.
\end{equation}

\begin{figure}[h!]
\begin{minipage}[t]{0.45\linewidth} 
\includegraphics[height=5.5cm]{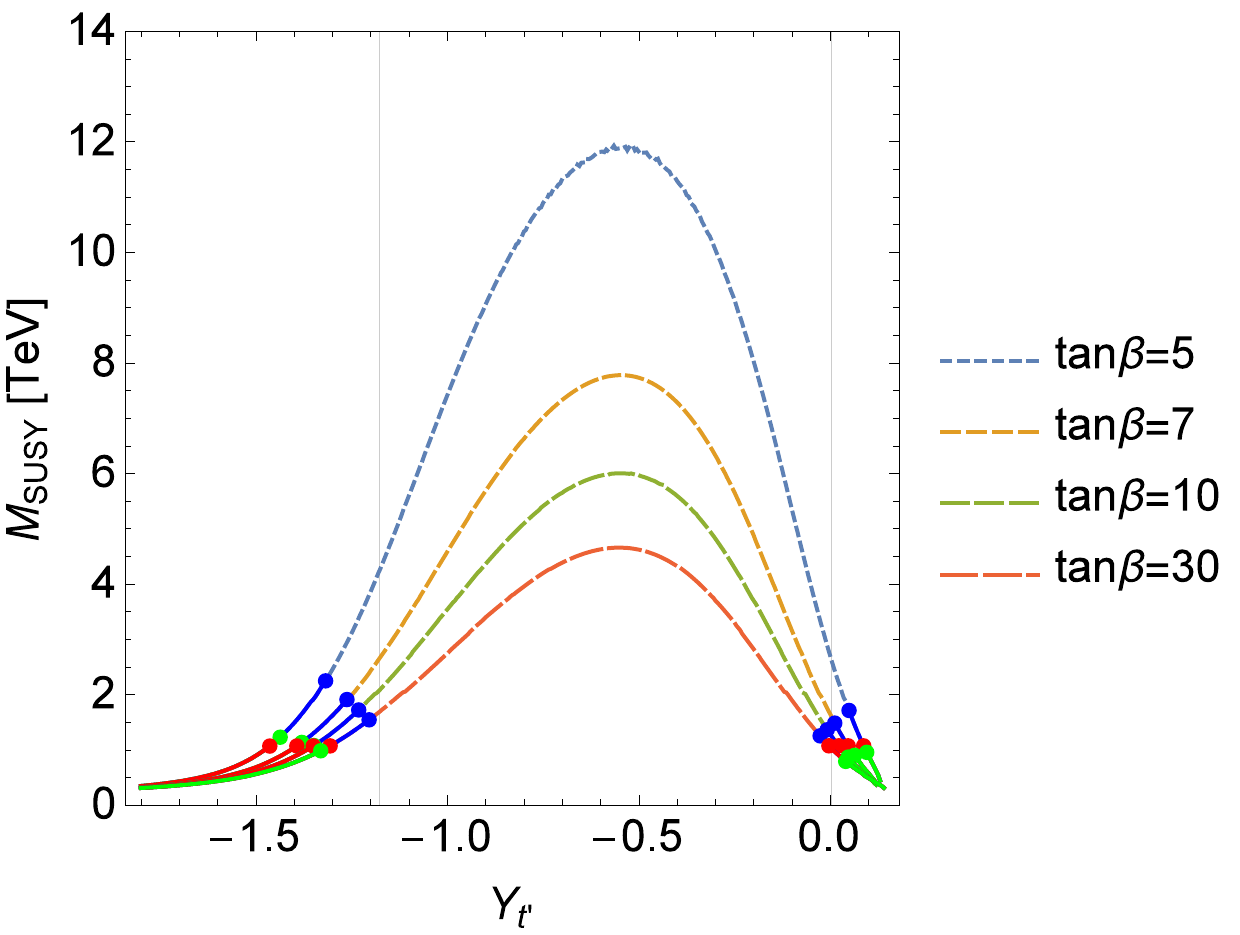} 
\end{minipage}
\begin{minipage}[t]{0.45\linewidth}
\includegraphics[height=5.5cm]{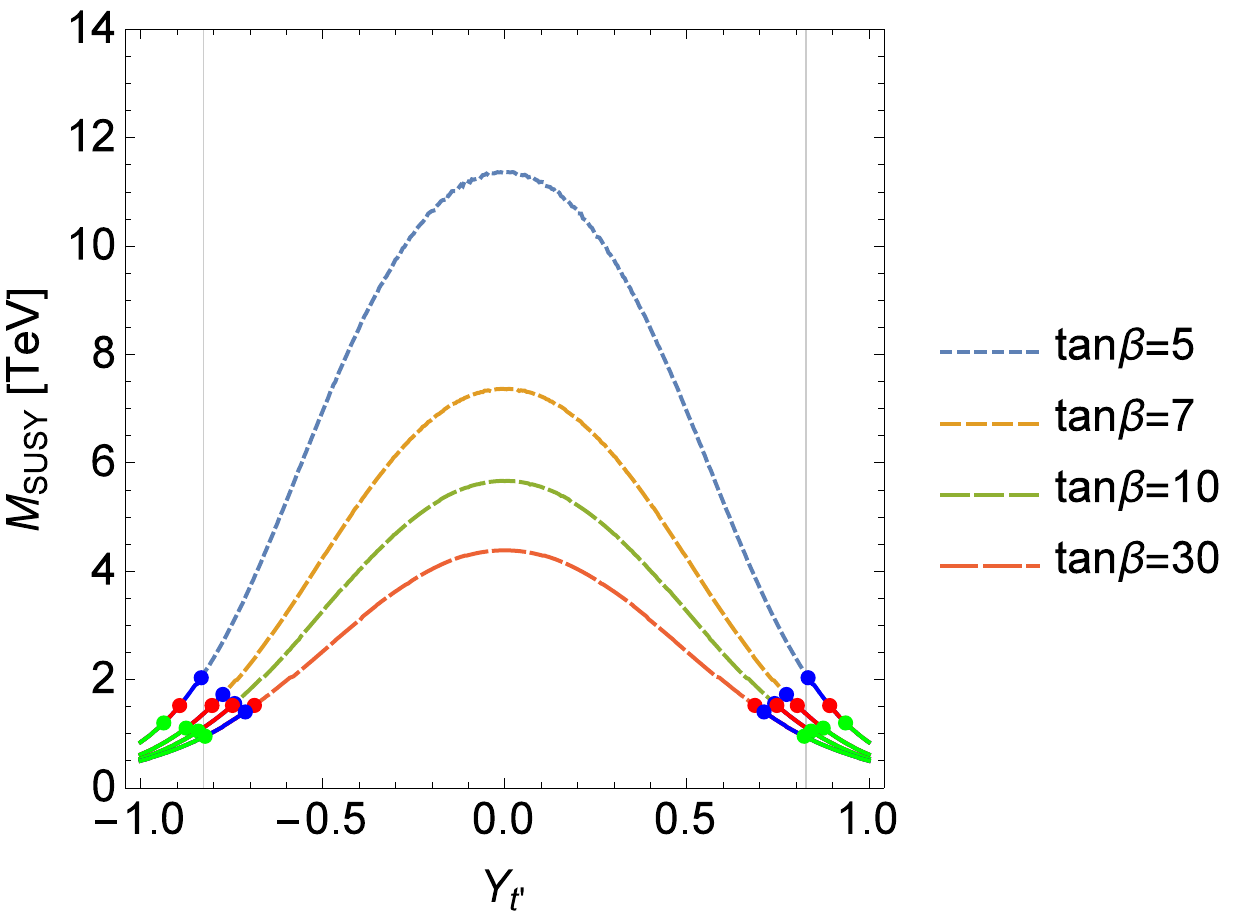}  
\end{minipage}
\begin{minipage}[t]{0.45\linewidth} 
\includegraphics[height=5.5cm]{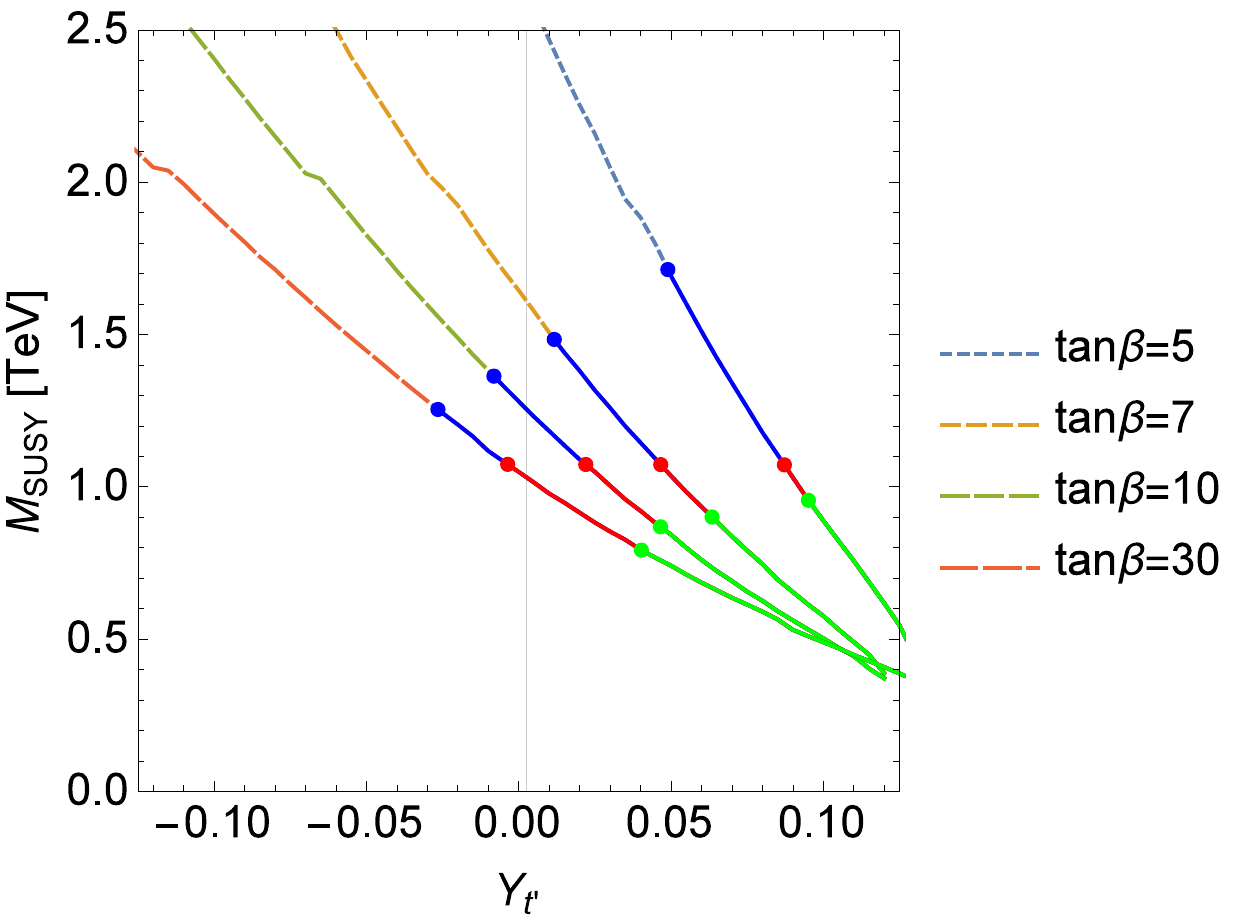} 
\end{minipage}
\begin{minipage}[t]{0.45\linewidth}
\includegraphics[height=5.5cm]{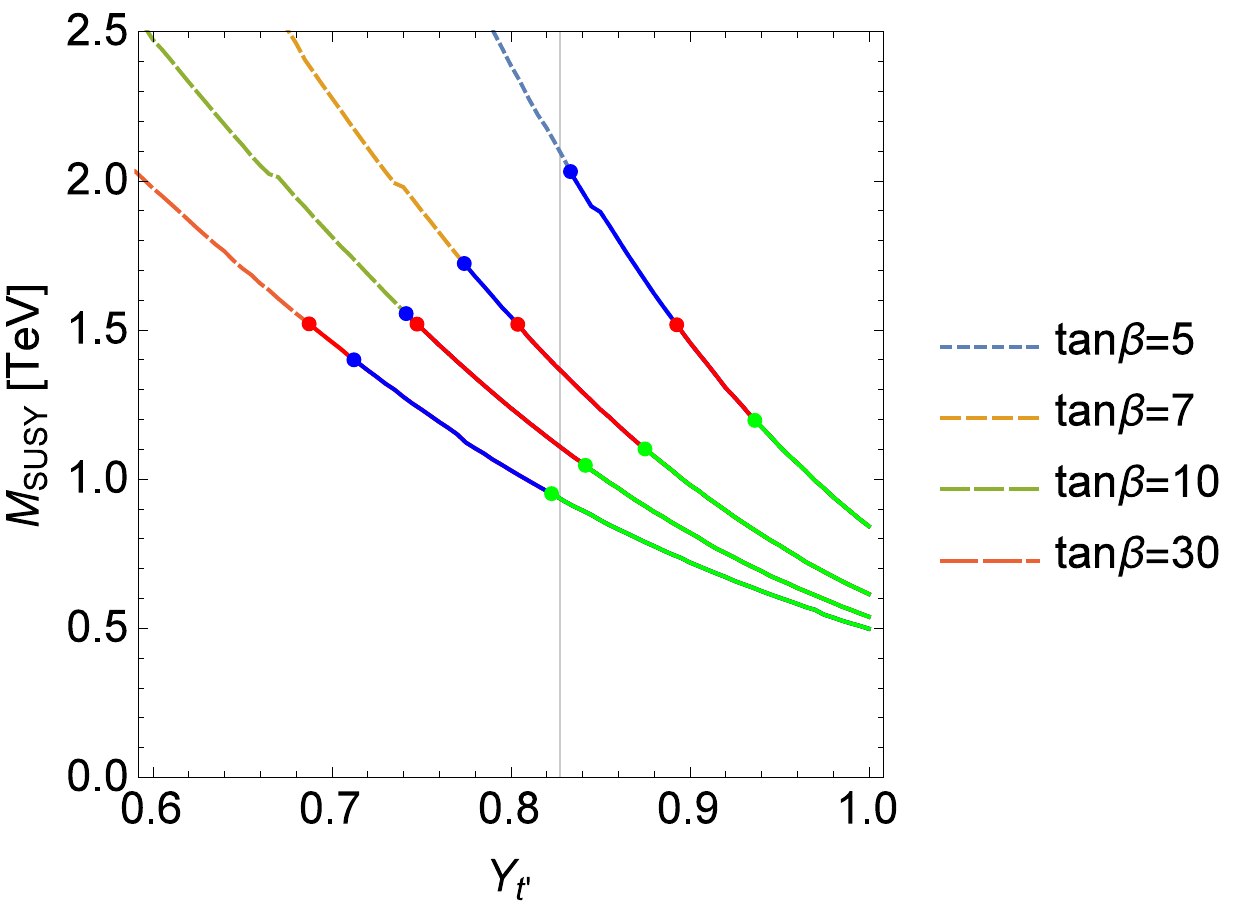}  
\end{minipage}
\caption{
Common superpartner mass $M_{SUSY}$ required  to obtain $m_h=\, $\mh as a function of $Y_{t'}$ for $m=M$ (left panel) and  $m=0$ (right panel).
Bottom row shows a zoom of the top row plots` lower right corners. MSSM values of $M_{SUSY}$ required  to obtain $m_h=\, $\mh corresponding to $~\tan{\beta}=5,7,10$ and $30$ are $M_{SUSY}=11.4 , 7.4 , 5.7$ and $4.4 \textrm{TeV}$.
Dashed lines are allowed by all considered constraints, while solid lines correspond to different exclusions which will be achievable in HL-LHC. The calculation of these bounds is explained in section \ref{Constraintssection}.
Dark blue regions may be excluded by measurement of the Higgs boson signal strength at $2\sigma$ significance. Dark green regions predict corrections to oblique parameters that may be excluded by future HL-LHC measurements at $2\sigma$ significance, and red regions may be excluded in the second LHC run by direct detection of the top partner.  Vertical lines show maximal $Y_{t'}$ allowing gauge coupling unification before the quasifixed point sets in. All parameters except $\tan{\beta}$ are fixed by assuming a single supersymmetry scale $M_{SUSY}$ and requiring correct top and Higgs physical masses $m_t=\, $\mt, $m_h=\, $\mh. 
\label{Msusy}
}
\end{figure}

\begin{figure}[ht]
\begin{minipage}[t]{0.45\linewidth} 
\centering
\includegraphics[height=6.1cm]{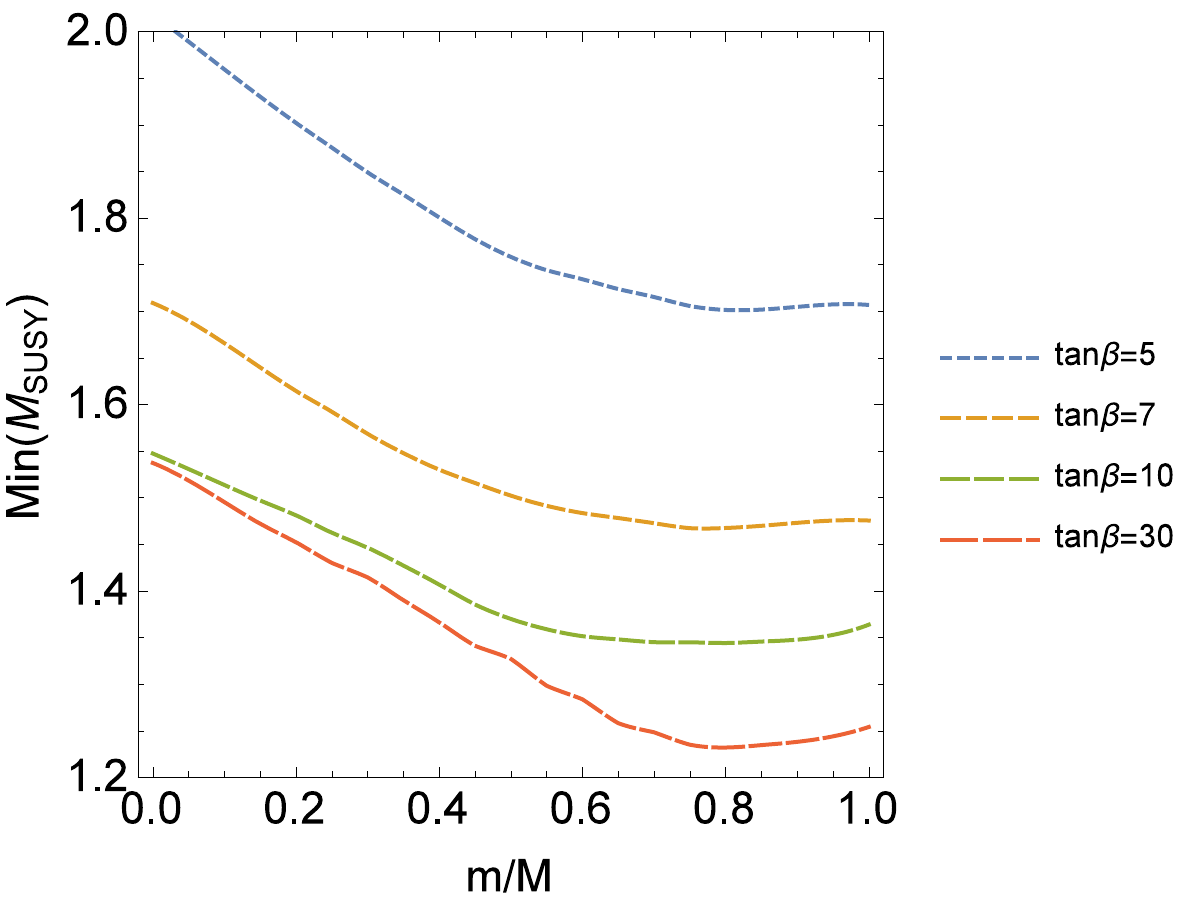} 
\end{minipage}
\hspace{0.5cm}
\begin{minipage}[t]{0.45\linewidth}
\centering 
\includegraphics[height=6.1cm]{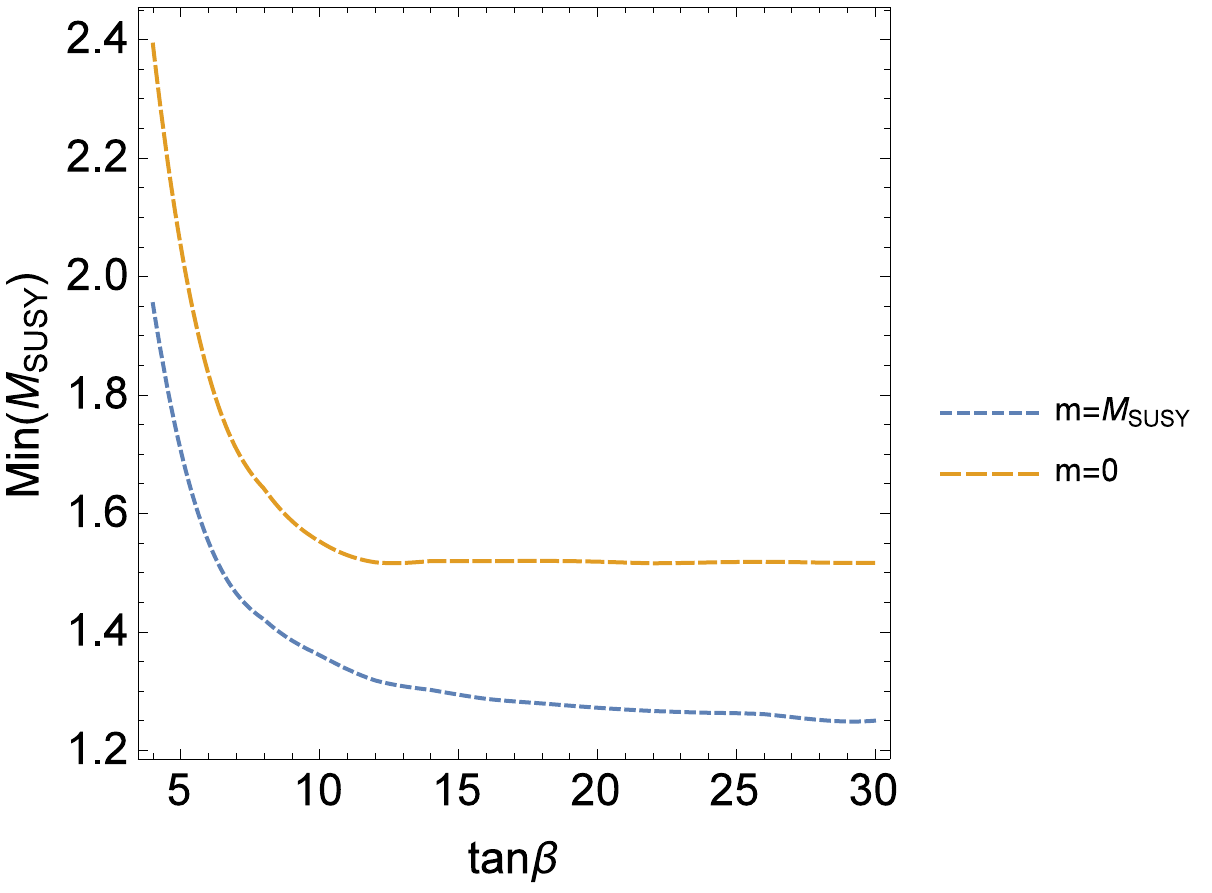}  
\end{minipage}
\caption{
Minimal value of $M_{SUSY}$ achievable without violating any of the above constraints as a function of $\frac{m}{M_{SUSY}}$ (left panel) and $\tan{\beta}$ (right panel). All other parameters are fixed by assuming a single supersymmetry scale $M_{SUSY}$ and requiring correct top and Higgs masses $m_t=\, $\mt, $m_h=\, $\mh.
\label{Msusy2}
}
\end{figure}

Figure~\ref{Msusy} shows the value of $M_{SUSY}$ needed to obtain $m_h=\, $\mh as a function of $Y_{t'}$ together with various constraints explained in the following section. Figure~\ref{Msusy2} shows the minimal value of $M_{SUSY}$ achievable without violating any of the experimental constraints. The smaller the value of $M_{SUSY}$ the more the vectorlike extension of the MSSM helps to ameliorate the little hierarchy problem. The MSSM values of $M_{SUSY}$ corresponding to $\tan \beta=5,7,10$ and $30$ are $M_{SUSY}=11.4 , 7.4 , 5.7$ and $4.4 \ \textrm{TeV}$, which means that in all presented cases we are able to achieve much lower $M_{SUSY}$ than required in the MSSM, without violating any of the constraints. 

Since the additional contribution to the Higgs mass from the vectorlike quark sector lowers the value of $M_{SUSY}$ needed to achieve the observed Higgs mass, it also increases the prospects of finding the correspondingly lower superpartner masses at subsequent runs of the LHC .

\subsection{RGE corrections}\label{RGEsection}

The introduction of additional states and additional Yukawa couplings to the MSSM causes the renormalization group flow trajectories of the couplings to be altered as the scale increases. In this section we discuss these effects and specify the implications and constraints they have on the unification of couplings and the possible development of Landau poles in the couplings.

In this analysis we have calculated two-loop renormalization group equations using SARAH \cite{Staub:2013tta}, and confirmed the results analytically using known results \cite{Martin:1993zk}. Very significant changes in the renormalization group trajectories come from new coefficients in the one-loop running of the gauge couplings,
\begin{equation}
\frac{d}{dt}g_i=\frac{1}{4\pi^2} \ b_i g_i^3 \quad \quad b_i= \left( \frac{41}{5},1,-2 \right).
\end{equation} 
These new equations predict the unification scale $M_U$ (defined here by $g_1(M_U)=g_2(M_U)$) to be significantly lower than in the MSSM. The new unification scale is not far above $10^{13}\ \textrm{GeV}$.

It is important to point out that unification at a scale around $10^{16} \ \textrm{GeV}$ can still easily be achieved by positing appropriate high-scale threshold corrections~\cite{Ellis:2015} or by adding vectorlike quarks so that together all vectorlike superfields form a complete representation of $SU(5)$. This can reestablish coupling constant unification without significant modifications to other bounds discussed in the following sections. 

However a more stringent constraint comes from the running of $Y_{t'}$ and its contribution to the running of $Y_t$. At one-loop order these contributions induce Landau poles in the Yukawa couplings' running when $Y_{t'}$ is sufficiently large --- at two-loop order $Y_t$ and $Y_{t'}$ develop a strongly coupled UV quasifixed point. The range of values of $Y_{t'}$ that allow gauge coupling unification before the UV quasifixed point sets in are $Y_{t'}\in (-1.775,0.002)$ for $m=M_{SUSY}$ and $Y_{t'}\in (-0.8275,0.8275)$ for $m=0$. These values are marked on the plots showing our results. However, since we do not consider a specific UV completion, it is not necessary to treat them as constraints. 

\section{Constraints}
\label{Constraintssection}
\subsection{Oblique parameter corrections}
\label{STUsection}

We calculate the $S$ and $T$ parameter~\cite{Peskin:1990zt}  contributions from the vectorlike quarks and their scalar superpartners using results from \cite{Martin:2009bg}, details are shown in Appendix \ref{STappendix}. To calculate MSSM contributions we use expressions from \cite{Martin:2004id} excluding corrections from stops and sbottoms which were already included in the vectorlike contribution calculation. We verified dominant corrections coming from new fermions with similar results from \cite{Lavoura:1992np}.

The currently allowed experimental values are $S=0.06\pm0.09$ and $T=0.1\pm0.07$ (assuming $U=0$) with correlation $0.91$ \cite{Baak:2014ora} (the correlation parameter is the tilt in the ellipse in the $S$-$T$ plane). Only minimally more stringent constraints can be achieved from LHC running at $\sqrt{s}=14\tev$ with high integrated luminosity $300 \, \textrm{fb}^{-1}$. Predicted future sensitivity values of $S=0.06\pm0.09$ and $T=0.1\pm0.06$ are taken from \cite{Baak:2013fwa}.

\begin{figure}[t]
\begin{minipage}[t]{0.45\linewidth} 
\centering
\includegraphics[height=6cm]{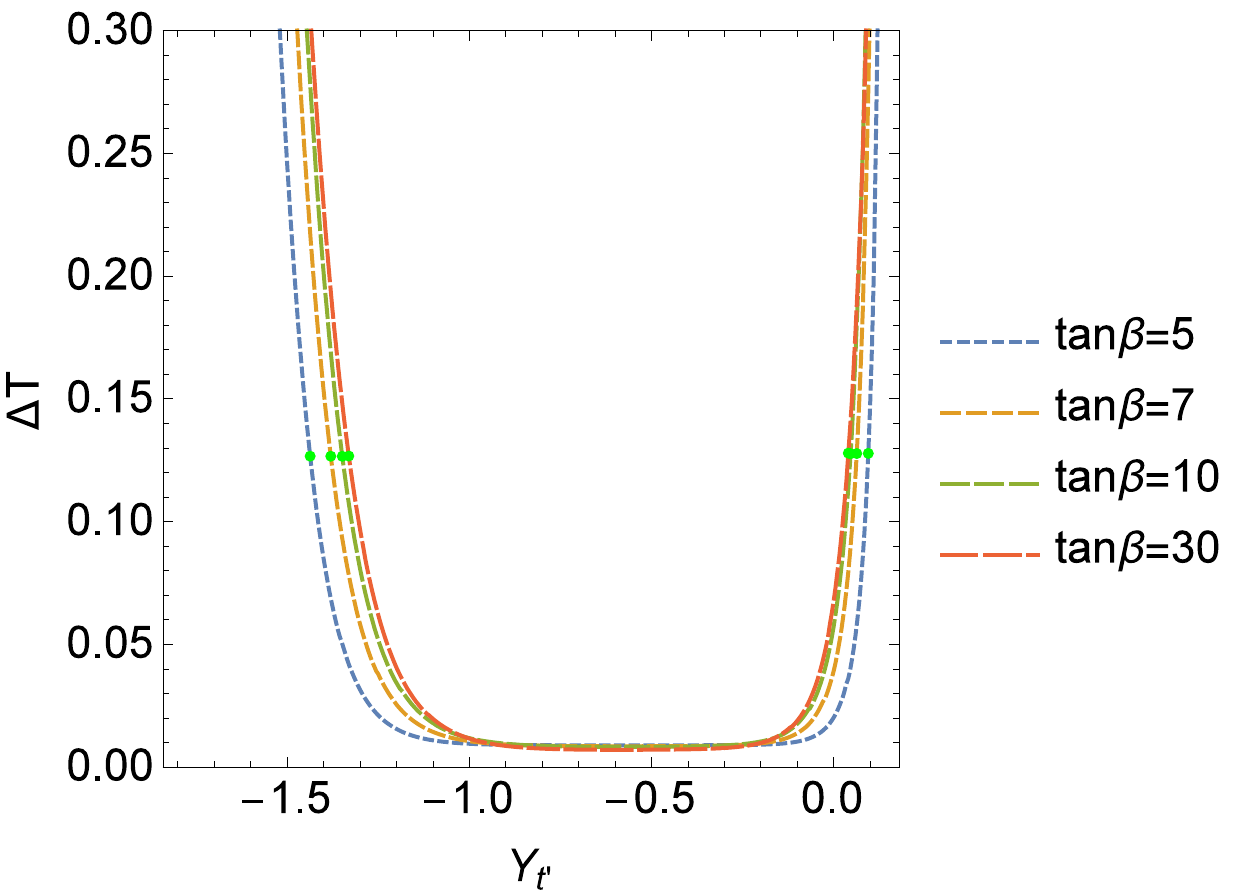} 
\end{minipage}
\hspace{0.5cm}
\begin{minipage}[t]{0.45\linewidth}
\centering 
\includegraphics[height=6cm]{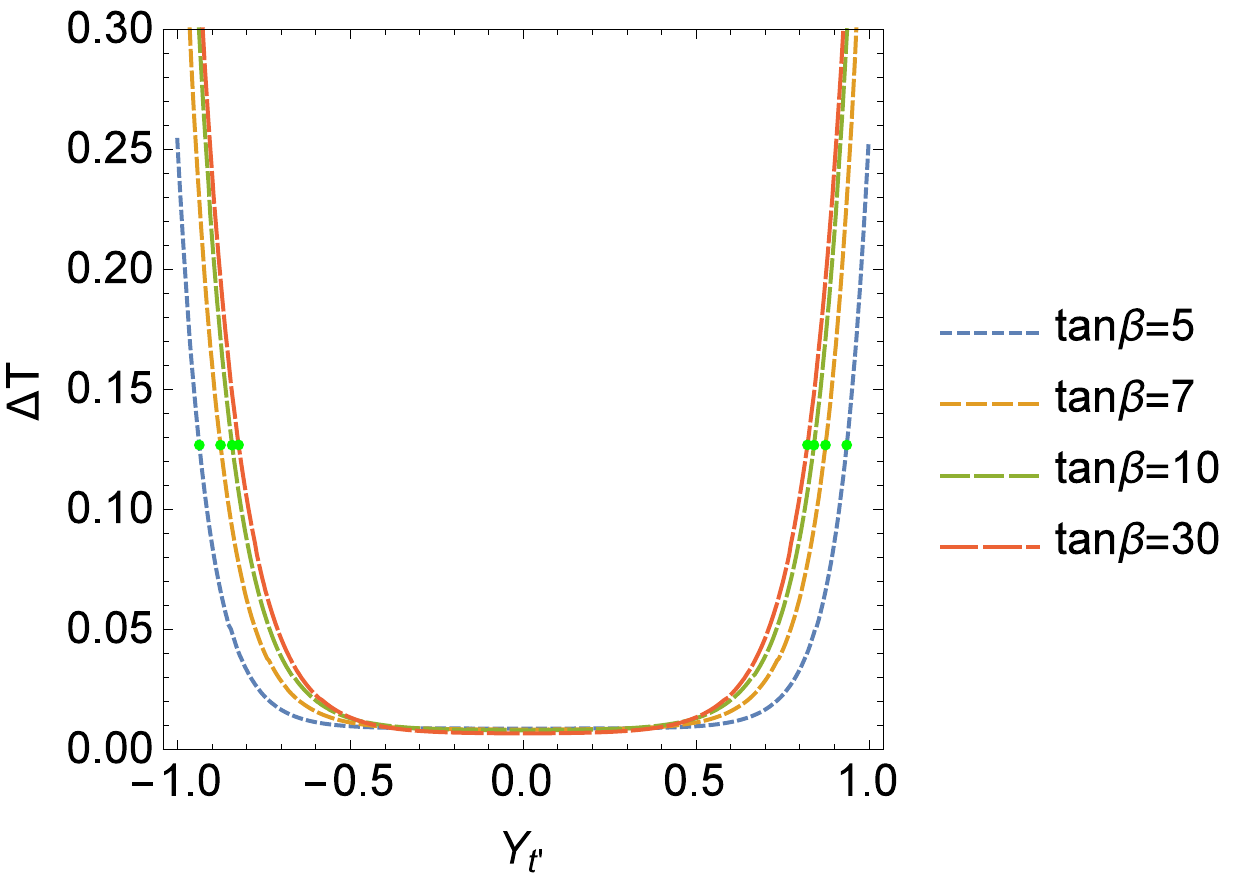}  
\end{minipage}
\caption{
Correction to the $T$ parameter as a function of $Y_{t'}$ for $m=M$ (left panel) and  $m=0$ (right panel). All values  satisfy $m_h=\, $\mh. Green points show values above which the results can be excluded at $2\sigma$ by future experimental constraints.
\label{STPlot}
}
\end{figure}

Figure~\ref{STPlot} shows resulting corrections to  the $T$ parameter as a function of $Y_{t'}$ together with points showing values above which the results can be excluded at $2\sigma$ by future experimental constraints. These points are very close to forming a vertical line because corrections to the $S$ parameter are very small for all interesting values of $Y_{t'}$. This is also the reason for which we do not include a plot of vectorlike corrections in the $S$- $T$ plane.  

\begin{figure}[h!]
\begin{minipage}[t]{0.45\linewidth} 
\centering 
\includegraphics[height=5.7cm]{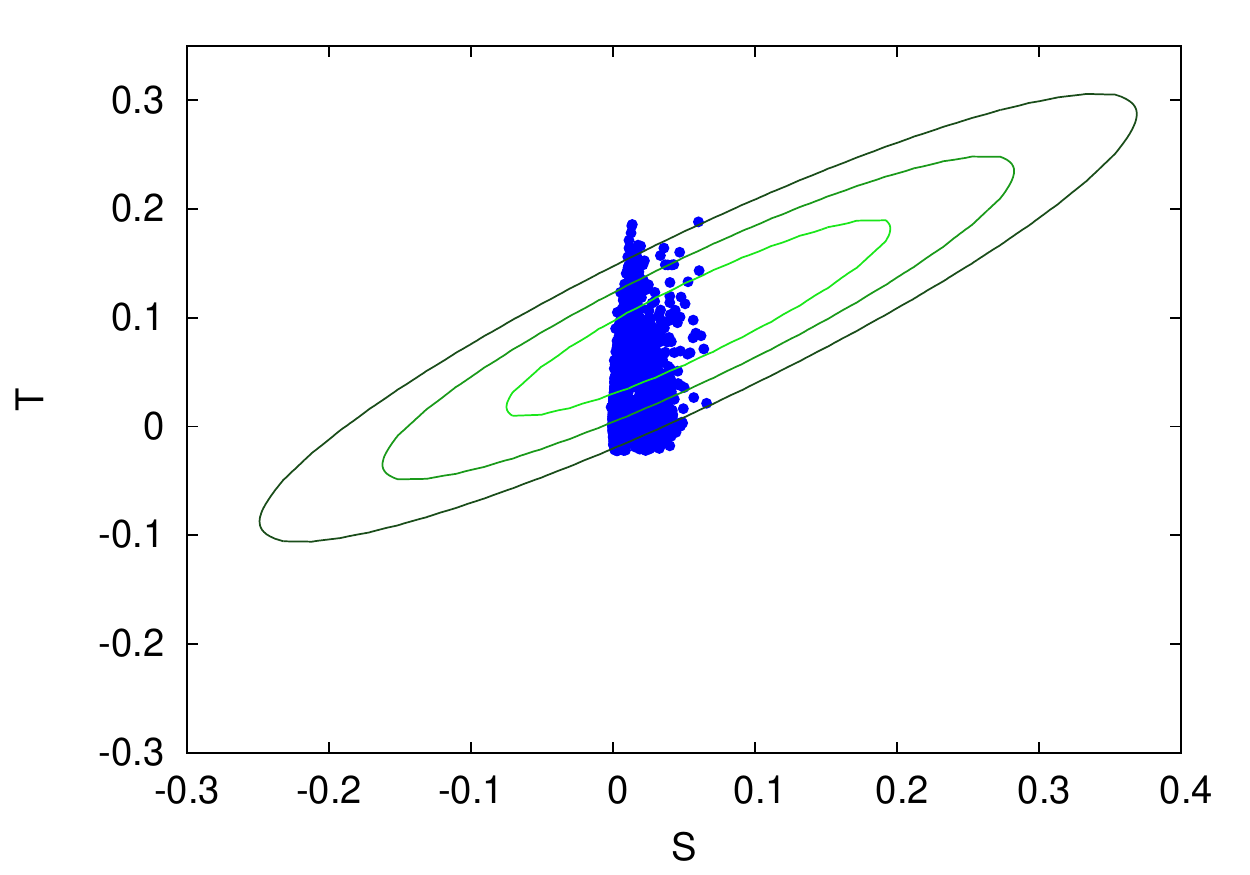}
\end{minipage}
\hspace{0.5cm}
\begin{minipage}[t]{0.45\linewidth}
\centering  
\includegraphics[height=5.7cm]{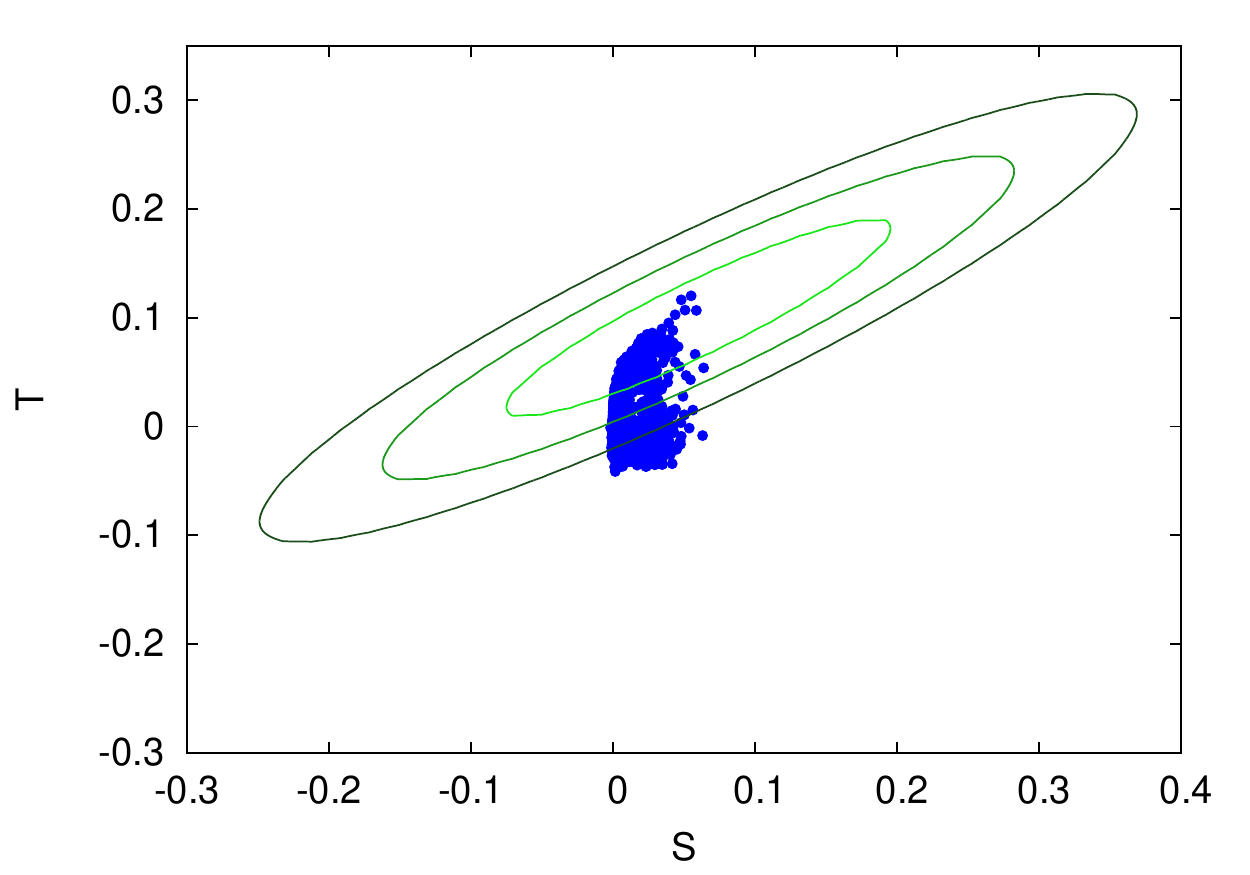}
\end{minipage}
\caption{
Oblique parameter corrections in $S - T$ plane coming from the MSSM (left panel) and the same results without stop and sbottom contribution (right panel), with a randomized spectrum of superpartner masses up to $3\tev$.
\label{MSSMST}
}
\end{figure}

Corrections from other superpartners are very small due to the simplified spectrum we chose. Figure~\ref{MSSMST} shows corrections coming from MSSM with and without the stops contribution from $100,000$ randomized spectra of masses up to $3\,\textrm{TeV}$. A more randomized spectrum is unlikely to produce points outside the the $S$ and $T$ exclusion ellipse. Most of the points would bring our results closer to the central values due to negative $T$ competing against large positive vectorlike quark corrections and positive $S$ contributions, which push our results towards the experimentally allowed ellipsis.

Superpartner corrections to electroweak precision observables are generally small because superpartners are largely decoupled even with current direct detection exclusions. However inclusion of a new quark can introduce unacceptably large corrections to the $T$ parameter if its mixing with the SM top is substantial.
Nevertheless, it is important to note that with currently available bounds, electroweak corrections are the most important  constraints on our model. However, as the energy and luminosity increase for HL-LHC the observables at play in the electroweak precision analysis do not improve substantially. Therefore, precision electroweak analysis constraints become relatively less important in time compared to direct detection probes of new states and especially compared to precision Higgs analysis, which is discussed in the next section.

\subsection{Higgs boson coupling corrections}\label{sec:Higgsprecisionsection}

Next we turn to calculation of Higgs boson branching ratios including the above modifications and new couplings to the top quark and its vectorlike partner.
We start by discussing the shifts in couplings of the MSSM compared to the SM and then compare with the case with extra vectorlike top states. In the MSSM, the Higgs couplings to up and down type quarks and vector bosons take the form \cite{MSSM Higgs,HiggsHuntersGuide}:
\begin{eqnarray}\label{MSSMcouplings}
c_u=\frac{g_u}{g_u^{\rm SM}}&=&\frac{\cos \alpha }{\sin \beta } \nonumber \\
c_d=\frac{g_d}{g_d^{\rm SM}}&=&\frac{-\sin \alpha }{\cos \beta } \\
c_V=\frac{g_V}{g_V^{\rm SM}}&=&\sin (\beta-\alpha),  \nonumber
\end{eqnarray} 
where $\alpha$ is the Higgs mixing angle and $\tan\beta=v_u/v_d$.

 Most experimentally important branching ratios have the same values as in the MSSM, which are obtained by multiplying the appropriate $c_i$ coefficients in front of the SM partial width exprressoins
\begin{eqnarray}
\Gamma(h\rightarrow b\bar{b}) &=& c_d^2 \Gamma^{\rm SM}(h\rightarrow b\bar{b}), \quad
\Gamma(h\rightarrow \tau \bar{\tau}) = c_d^2 \Gamma^{\rm SM}(h\rightarrow \tau \bar{\tau}), \nonumber \\ \Gamma(h\rightarrow \mu \bar{\mu}) &=& c_d^2 \Gamma^{\rm SM}(h\rightarrow \mu \bar{\mu}) ,\quad
\Gamma(h\rightarrow c\bar{c}) = c_u^2 \Gamma^{\rm SM}(h\rightarrow c\bar{c}), \\
\Gamma(h\rightarrow WW) &=& c_V^2 \Gamma^{\rm SM}(h\rightarrow WW), \quad
\Gamma(h\rightarrow ZZ) = c_V^2 \Gamma^{\rm SM}(h\rightarrow ZZ). \nonumber
\end{eqnarray}
The remaining important branching ratios are loop induced and are modified due to modified top couplings and new particles in the loops. We will express these branching ratios as
\begin{equation}
\Gamma (h\rightarrow X)= \frac{\left|\mathcal{A}_X\right|^2}{\left|\mathcal{A}_X^{\rm SM} \right|^2} \Gamma (h\rightarrow X)^{\rm SM}.
\end{equation}
In the following $N_c=3$ and loop functions $F$, $I$ and $A$, as well as coefficients $\tau$, are defined in \cite{MSSM Higgs}. Charges and third components of isospin for fields used in the following equations are shown in Table~\ref{eT3tab}, while modifications of the top and top prime couplings to the Higgs bosons are given by 
\begin{equation}
g_{ h t_i \bar{t}_i}=\frac{Y_t L_{i 1}R_{i 1}+Y_{t'}L_{i 1}R_{i 2}}{Y_t^{\rm MSSM}},
\end{equation}
where $L$ and $R$ are fermion mixing matrices defined in \eqref{LRmatrixdef}. $\mathcal{A}_{X}^{SUSY}$ are sums of the contributions of superpartners which we neglect since they have very small couplings $g \approx \frac{m^2_Z}{M_{SUSY}^2}$.

For branching ratio to two gluons we have,
\begin{eqnarray}
\mathcal{A}_{g g} &=&
c_d \sum\limits_{i=d,s,b}F_{\frac{1}{2}}(\tau_i)
+c_u \sum\limits_{i=u,c}F_{\frac{1}{2}}(\tau_i)
+c_u \sum\limits_{i=1}^{2}g_{h t_i  \bar{t}_i} F_{\frac{1}{2}}(\tau_{t_i})
+\mathcal{A}_{g g}^{SUSY}, \\
\mathcal{A}_{g g}^{\rm SM} &=&
 \sum\limits_{i=d,s,b}F_{\frac{1}{2}}(\tau_i)+
 \sum\limits_{i=u,c,t}F_{\frac{1}{2}}(\tau_i). \nonumber
\end{eqnarray} 
Similarly for the branching ratio to two photons we have,
 \begin{eqnarray}
\mathcal{A}_{\gamma \gamma} &=&
 c_V F_1(\tau_W)
+c_d e_e^2\sum\limits_{i=e,\mu,\tau}F_{\frac{1}{2}}(\tau_i)
+c_d N_c e_d^2\sum\limits_{i=d,s,b}F_{\frac{1}{2}}(\tau_i)
+c_u N_c e_u^2 \sum\limits_{i=u,c}F_{\frac{1}{2}}(\tau_i)
\nonumber \\ &+&
c_u N_c e_u^2 \sum\limits_{i=1}^{2}g_{h t_i  \bar{t}_i} F_{\frac{1}{2}}(\tau_{t_i})
+\mathcal{A}_{\gamma \gamma}^{\rm SUSY} \\
\mathcal{A}_{\gamma \gamma}^{\rm SM} &=&
F_1(\tau_W)
+e_e^2\sum\limits_{i=e,\mu,\tau}F_{\frac{1}{2}}(\tau_i)
+N_c e_d^2 \sum\limits_{i=d,s,b}F_{\frac{1}{2}}(\tau_i)
+N_c e_u^2 \sum\limits_{i=u,c,t}F_{\frac{1}{2}}(\tau_i)
. \nonumber
\end{eqnarray} 
Lastly for branching ratio of Higgs to a photon and $Z$ boson we obtain
\begin{eqnarray}
\mathcal{A}_{Z \gamma} &=&
c_d e_e v_e \sum\limits_{i=e, \mu, \tau}A_{\frac{1}{2}}(\tau_i,\lambda_i)
+
c_d N_c e_d v_d \sum\limits_{i=d,s,b}A_{\frac{1}{2}}(\tau_i,\lambda_i)
+
c_u N_c e_u v_u \sum\limits_{i=u,c}A_{\frac{1}{2}}(\tau_i,\lambda_i)
\nonumber \\ &+&
c_u N_c e_u \sum\limits_{i=1}^{2} v_{t_i} g_{h t_i  \bar{t}_i} A_{\frac{1}{2}}(\tau_{t_i},\lambda_{t_i})
+
 c_V A_{1}(\tau_W,\lambda_W) + \mathcal{A}_{Z \gamma}^{SUSY}
   \\
\mathcal{A}_{Z \gamma}^{\rm SM} &=&
e_e v_e \sum\limits_{i=e, \mu, \tau}A_{\frac{1}{2}}(\tau_i,\lambda_i)
+
N_c e_d^2 \sum\limits_{i=d,s,b}A_{\frac{1}{2}}(\tau_i,\lambda_i)
+
N_c e_u^2 \sum\limits_{i=u,c,t}A_{\frac{1}{2}}(\tau_i,\lambda_i)
+
A_1(\tau_W,\lambda_W)
, \nonumber
\end{eqnarray}
where $v_f=(2T_3^f -4 e_f s_W^2)/(s_W c_W)$, $s_W=\sin \theta_W$ and $c_W=\cos \theta_W$.
\begin{table}[t]
\centering
\begin{tabular}{c|c c c c c}
$f$ & $t_i$ & $u$ & $d$ & $e$&
\\ \hline 
$e_f$ & $\frac{2}{3}$& $\frac{2}{3}$& $-\frac{1}{3}$& $-1$&
 \\
$T_3^f$ & $\frac{1}{2} L_{i 1}$ & $\frac{1}{2}$& $-\frac{1}{2}$& $-\frac{1}{2}$& 
\end{tabular}
\caption{Charges and effective third isospin components. The mixing matrix $L$ is defined in \eqref{LRmatrixdef}.
}\label{eT3tab}
\end{table}

The branching ratios are given by
\begin{equation}
B(h \rightarrow X)=\frac{\Gamma_X}{\sum\limits_i \Gamma_i}
\end{equation}
with the sum running over all decay channels computed in this section.
We approximate the resulting signal strength modification by including only the gluon fusion production channel, which at leading order gives
\begin{eqnarray}
\Delta \mu_X &=& \frac{\sigma B(h \rightarrow X)-\sigma^{\rm SM} B^{\rm SM}(h \rightarrow X)}{ \sigma^{\rm SM} B^{\rm SM}(h \rightarrow X)}=
 \frac{\sigma B(h \rightarrow X)}{ \sigma^{\rm SM} B^{\rm SM}(h \rightarrow X)}-1
  \\ \nonumber & \approx &
\frac{\sigma( g g \rightarrow h)}{\sigma^{\rm SM}( g g \rightarrow h)} \frac{Br(h \rightarrow X)}{B^{\rm SM}(h \rightarrow X)}-1
\approx 
\frac{\Gamma(h \rightarrow g g )}{\Gamma^{\rm SM}(h \rightarrow g g)} \frac{B(h \rightarrow X)}{B^{\rm SM}(h \rightarrow X)}-1.
\end{eqnarray}

We confront these results with future experimental bounds as predicted by the CMS Collaboration \cite{CMS:2013xfa} shown in Table~\ref{mutab}.
SM values of the branching ratios were taken from \cite{Almeida:2013jfa}.
The resulting signal strength modifications are dominated by the increased $gg\rightarrow H$  production cross section compared to the SM and even MSSM. In our model all signal strengths grow rapidly when the mixing with the vectorlike state is increased.
The most important exclusion limit comes from the $H \rightarrow WW$ signal. The high sensitivity in this channel is due to the onset of high statistics and high accuracy in the measurement of this channel at the HL-LHC. This can be compared  to $H\rightarrow \gamma \gamma$ which is not as useful due to smaller modifications of its total $\sigma\cdot {\rm B}$ rate. There is a partial cancellation of vectorlike top contribution in the $\sigma\cdot {\rm B}$ product. The second best exclusion channel is $H\rightarrow ZZ$ with slightly worse experimental accuracy.
The increased experimental sensitivities at HL-LHC leads to the conclusion that the first evidence for vectorlike quarks in this context of natural supersymmetry would likely come from deviations found in precision Higgs observables.

\begin{table}[t]
\centering
\begin{tabular}{| c | c | c | c | c |}
\hline
$\Delta \mu_{\gamma \gamma}$ &$\Delta \mu_{b b}$ & $\Delta \mu_{\tau \tau}$ & $\Delta \mu_{WW}$ & $\Delta \mu_{ZZ}$   \\ \hline
$0.06$ &  $0.11$ & $0.08$ & $0.06$ & $0.07$ \\
\hline
\end{tabular}
\caption{Higgs signal strength future experimental sensitivities at $1\sigma$ significance from CMS \cite{CMS:2013xfa}
}\label{mutab}
\end{table}

\subsection{Direct Detection}\label{DirectDetection}

The best source for the direct mass bound for the new vectorlike states are dedicated analyses by the ATLAS and CMS collaborations at LHC. In particular,  the recent CMS analysis \cite{Chatrchyan:2013uxa} of $t'$ decaying in three channels $t' \rightarrow bW,tZ,tH$ without assumptions on the branching ratios, has current mass limits between $687 \gev$ and $782 \gev$.

A similar analysis of decay to the same final states in future colliders was performed in \cite{Bhattacharya:2013iea}. The authors predict mass ranges in which $t'$ could be discovered or excluded for different energies and integrated luminosities.
We use their exclusion limit (at $95\%$ C.L.) for vectorlike top partner achievable in LHC at $\sqrt{s}=14 \tev$ with integrated luminosity $300 \, \textrm{fb}^{-1}$, namely $m_{t'}<1525 \gev$.

\begin{figure}[t]
\begin{minipage}[t]{0.45\linewidth} 
\centering 
\includegraphics[height=6cm]{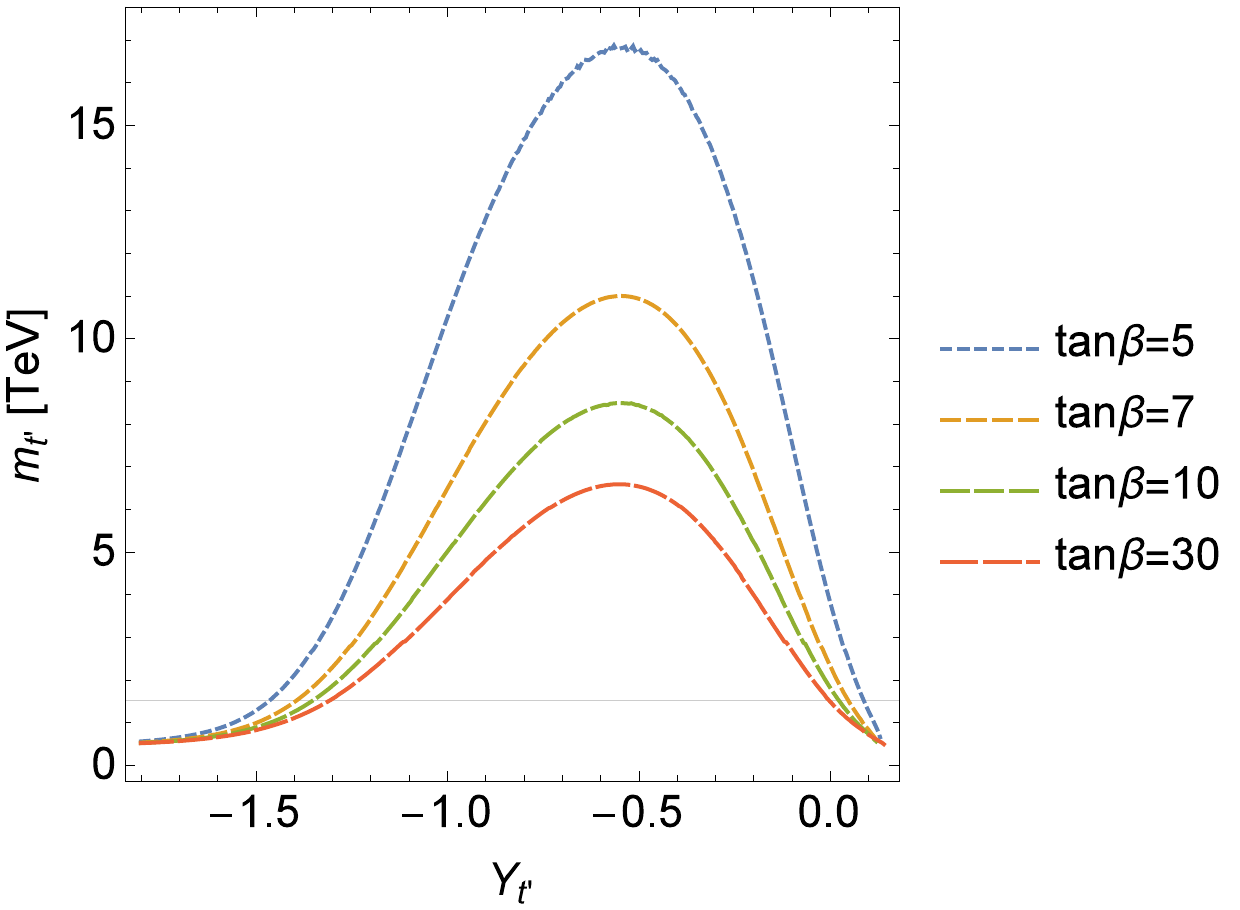}
\end{minipage}
\hspace{0.5cm}
\begin{minipage}[t]{0.45\linewidth}
\centering  
\includegraphics[height=6cm]{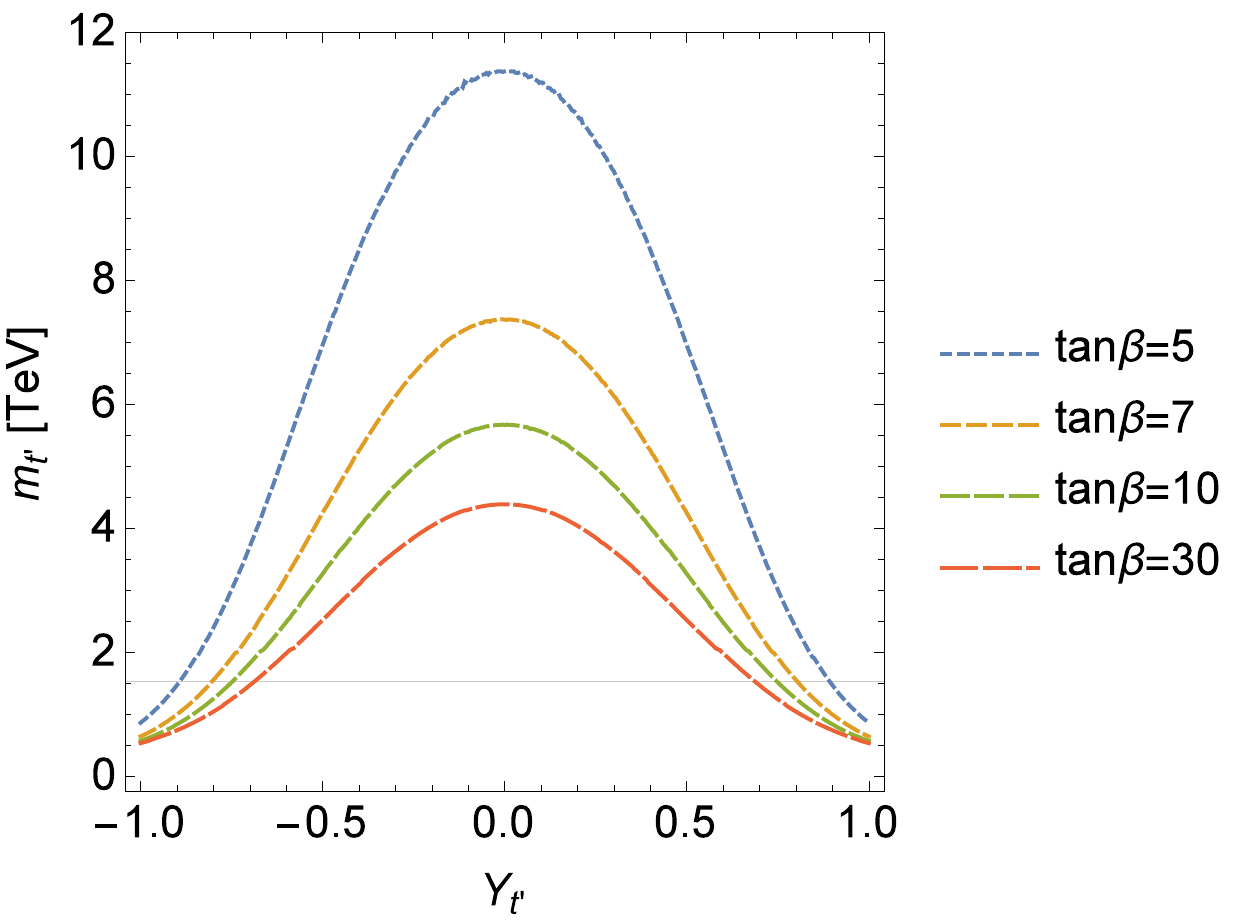}
\end{minipage}
\caption{
Vectorlike top partner mass for which $m_h=\, $\mh as a function of $Y_{t'}$ for $m=M$ (left panel) and  $m=0$ (right panel). Horizontal line corresponds to the future experimental bound.
\label{vectorlikemass}
}
\end{figure}

Figure~\ref{vectorlikemass} shows the vectorlike top partner mass needed to achieve $m_h=\, $\mh as a function of $Y_{t'}$. The right-hand side plot is very similar to Figure~\ref{Msusy} because, as expected, the mass of the vectorlike top is close to $M_{SUSY}$, while in the left-hand side plot the mass is significantly enhanced due to large mixing.

\begin{figure}[t]
\centering 
\includegraphics[height=6cm]{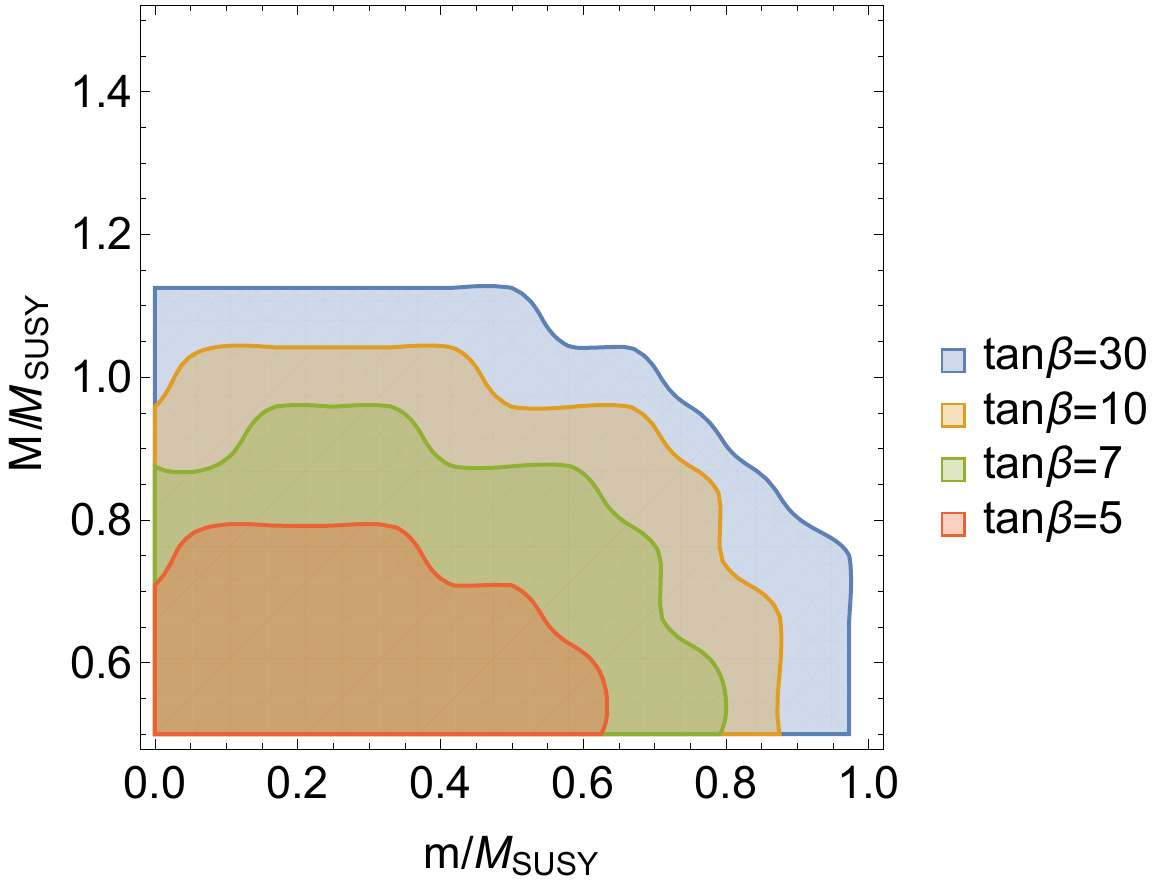}
\caption{
Region of vectorlike mass and mixing parameter space, where direct detection is the strongest constraint. The unmarked regions corresponds to precision Higgs measurements being the strongest constraint.
\label{Mmplot}
}
\end{figure}

It is important to point out here that direct detection is crucially dependent on the mass of the additional quark, while all previously discussed constraints were more dependent on its mixing with already observed states. Consequently the interplay between constraints described in this section and those of the previous two depends on the mixing, which is a consequence of our choice of spectrum parameters. This is why we include both small ($m=0$) and maximal ($m=M_{SUSY}$) mixing scenarios in our analysis. Direct detection bounds turn out to be very important for our model. And in fact this probe proves to be the strongest for the part of parameter space corresponding to large $\tan \beta$, unless the mixing is sufficiently large ($m \approx M_{SUSY}$). Otherwise precision Higgs analysis will be a more powerful probe as shown in Figure~\ref{Mmplot}.   
\section{Conclusions}

In summary we analyzed a single vectorlike top partner model, which is the simplest vectorlike extension of the MSSM that can significantly help with the little hierarchy problem. We calculated and compared different experimental constraints the model will face after $300\,\textrm{fb}^{-1}$ of data are gathered at the HL-LHC.
Our key result is that the most constraining of the discussed bounds is modification of the Higgs boson properties. An exception to that is the case of large $\tan \beta$ and small mixing  where the direct detection probes of the heavy vectorlike states at the collider are slightly more stringent. 

After including all the constraints achievable at the HL-LHC,  the resulting $M_{SUSY}$ can still be as low as $1.2$ to $2.4\,\textrm{TeV}$ for the simplest possible supersymmetry spectrum. 
These results are $3$ to $5$ times smaller  compared to what otherwise would be allowed in the MSSM. 
Thus even a very simple vectorlike quark extension can greatly reduce the little hierarchy problem of the MSSM, and careful measurements of Higgs boson observables would likely give first evidence of this scenario.

\vspace{0.25cm}
\noindent
{\it Acknowledgements:}
This work was supported by the Foundation for
Polish Science International PhD Projects Programme co-financed by the EU
European Regional Development Fund.
This work was partially supported by National Science Centre, Poland 
under research grants DEC-2012/04/A/ST2/00099 and DEC-2014/13/N/ST2/02712. JDW is supported
in part by DOE under grant DE-SC0011719.

\vspace*{0.3cm}
\appendix 
\section{Oblique parameter corrections}\label{STappendix}
The Peskin-Takeuchi precision electroweak parameters 
\cite{Peskin:1990zt} $S$ and $T$ are
defined in terms of electroweak vector boson self-energies as
\begin{eqnarray}
\frac{\alpha S}{4 s_W^2 c_W^2} &=& \left[
\Pi_{ZZ}(M_Z^2) - \Pi_{ZZ}(0) 
- \frac{c_{2W}}{c_W s_W} \Pi_{Z\gamma}(M_Z^2) - \Pi_{\gamma\gamma}(M_Z^2)
\right] /M_Z^2
,
\\
\alpha T &=& \Pi_{WW}(0)/M_W^2 - \Pi_{ZZ}(0)/M_Z^2
.
\end{eqnarray}
The one-loop functions $G(x)$, $H(x,y)$, 
$B(x,y)$, and $F(x,y)$ have been defined in ref.~\cite{Martin:2004id}. 
Particle names stand for the squared mass of the particles 
when they appear as an argument of these functions.

Contributions from $t'$ to the electroweak vector boson
self-energies are:
\begin{eqnarray}
\Delta \Pi_{\gamma\gamma} &=& 
-\frac{N_c}{16 \pi^2} 2 g^2 s_W^2 \left[ 
e_u^2  G(M_{t_2}) \right], \nonumber
\\
\Delta \Pi_{Z\gamma} &=& 
-\frac{N_c}{16 \pi^2} g s_W \left[ 
e_u \sum_{i=1,2} (g^{Z}_{t_i t_i^{\dagger}} - 
g^{Z}_{\bar t_i \bar t_i^{\dagger}}) G(t_i)  
\right]- \Delta \Pi_{Z\gamma}^{\rm SM}
,
\\
\Delta \Pi_{ZZ} &=& 
-\frac{N_c}{16 \pi^2} \left[ 
\sum_{i,j=1}^2 (|g^{Z}_{t_i t_j^{\dagger}}|^2 + 
|g^{Z}_{\bar t_i \bar t_j^{\dagger}}|^2) H(t_i, t_j) 
-4 {\rm Re}(
g^{Z}_{t_i t_j^{\dagger}} g^{Z}_{\bar t_i \bar t_j^{\dagger}}) 
m_{t_i} m_{t_j} B(t_i, t_j)
\right]-\Delta \Pi_{ZZ}^{\rm SM}
 \nonumber
,
\\
\Delta \Pi_{WW} &=& 
-\frac{N_c}{16 \pi^2} \sum\limits_{i=1}^{2} \left[
(|g^{W}_{t_i b^{\dagger}}|^2 
) H(b, t_i) 
\right]-\Delta \Pi_{WW}^{\rm SM} 
\nonumber,
\end{eqnarray}
where $N_c = 3$, $e_u = 2/3$, $e_d = -1/3$ and SM contributions are similar to those above with couplings in which $L_{11}=1$ is the only nonzero element of the mixing matrix. 
The massive vector boson couplings with quarks are
\begin{eqnarray}
g^{Z}_{t_i t_j^{\dagger}} &=& 
\frac{g}{c_W} \left (\frac{1}{2} L_{i1}^* L_{j1} 
- e_u s_W^2 \delta_{ij}\right ), \quad \quad 
g^{Z}_{\bar t_i \bar t_j^{\dagger}} = \frac{g}{c_W} \left ( 
 e_u s_W^2 \delta_{ij}\right ), \nonumber \\
g^{W}_{ t_i  b^{\dagger}} &=& \frac{g}{\sqrt{2}} L_{i1}^*,
\end{eqnarray}
where $L$ is the fermion mixing matrix defined in \eqref{LRmatrixdef}.

The up-type scalar mass matrix \eqref{scalarmass} is diagonalized by the unitary matrix $U$:
\begin{equation}
U {\mathbf M}^2_S U^{\dagger} = {\rm diag}(m^2_{\tilde{t}_1},m^2_{\tilde{t}_2},m^2_{\tilde{t}_3},m^2_{\tilde{t}_4})
,
\end{equation}
while the MSSM sbottom mass matrix ${\mathbf M}^2_D$
is diagonalized by the unitary matrix $D$
\begin{equation}
D {\mathbf M}^2_D D^{\dagger} = {\rm diag}(m^2_{\tilde{b}_1},m^2_{\tilde{b}_2})
.
\end{equation}
Contributions from third family squarks to the electroweak vector boson
self-energies are
\begin{eqnarray}
\Delta \Pi_{\gamma\gamma} &=& 
\frac{N_c}{16 \pi^2} g^2 s_W^2 \left [
       e_u^2 \sum_{i=1}^4 F(\tilde t_i, \tilde t_i)
      +e_d^2 \sum_{i=1}^2 F(\tilde b_i, \tilde b_i)
\right ], \nonumber
\\
\Delta \Pi_{Z\gamma} &=& 
\frac{N_c}{16 \pi^2} g s_W \left [ 
 e_u \sum_{i=1}^4 g^{Z}_{\tilde t_i \tilde t_i^{*}} 
 F(\tilde t_i, \tilde t_i)
+e_d \sum_{i=1}^2 g^{Z}_{\tilde b_i \tilde b_i^{*}} 
 F(\tilde b_i, \tilde b_i)
\right ],
\\
\Delta \Pi_{ZZ} &=& 
\frac{N_c}{16 \pi^2} \left [ 
 \sum_{i,j=1}^4 |g^{Z}_{\tilde t_i \tilde t_j^{*}} |^2  F(\tilde t_i, \tilde t_j) 
+\sum_{i,j=1}^2 |g^{Z}_{\tilde b_i \tilde b_j^{*}} |^2  F(\tilde b_i, \tilde b_j)
\right ]
, \nonumber
\\ \nonumber
\Delta \Pi_{WW} &=& 
\frac{N_c}{16 \pi^2} \sum_{i=1}^2 \sum_{j=1}^4
|g^{W}_{\tilde b_i \tilde t_j^{*}}|^2 F(\tilde b_i, \tilde t_j),
\end{eqnarray}
where the vector boson couplings with the squarks are:
\begin{eqnarray}
g^{Z}_{\tilde t_i \tilde t_j^{*}} &=& \frac{g}{c_W} \left (
\frac{1}{2} (U_{i1}^* U_{j1}) 
- e_u s_W^2 \delta_{ij}\right )
, \nonumber
\\ 
\qquad\qquad
g^{Z}_{\tilde b_i \tilde b_j^{*}} &=&  \frac{g}{c_W} \left (
-\frac{1}{2}(D_{i1}^* D_{j1}) - e_d s_W^2\delta_{ij} \right )
, 
\\
g^{W}_{\tilde b_i \tilde t_j^{*}} &=& 
\frac{g}{\sqrt{2}} ( D_{i1}^* U_{j1} ). \nonumber
\end{eqnarray}



\begin{thebibliography}{99}

\bibitem{Moroi:1991mg}
  T.~Moroi and Y.~Okada,
  Mod.\ Phys.\ Lett.\  A {\bf 7}, 187 (1992).

\bibitem{Moroi:1992zk}
  T.~Moroi and Y.~Okada,
  Phys.\ Lett.\  B {\bf 295}, 73 (1992).

\bibitem{Babu:2004xg}
  K.S.~Babu, I.~Gogoladze and C.~Kolda,
  ``Perturbative unification and Higgs boson mass bounds,''
  [hep-ph/0410085].

\bibitem{Babu:2008ge}
  K.S.~Babu, I.~Gogoladze, M.U.~Rehman and Q.~Shafi,
  Phys.\ Rev.\  D {\bf 78}, 055017 (2008)
  [hep-ph/0807.3055].

\bibitem{Martin:2009bg} 
  S.P.~Martin,
  Phys.\ Rev.\ D {\bf 81}, 035004 (2010)
  [0910.2732 [hep-ph]].

\bibitem{Graham:2009gy} 
  P.W.~Graham, A.~Ismail, S.~Rajendran and P.~Saraswat,
  Phys.\ Rev.\ D {\bf 81}, 055016 (2010)
  [0910.3020 [hep-ph]].

\bibitem{Martin:2010dc} 
  S.P.~Martin,
  Phys.\ Rev.\ D {\bf 82}, 055019 (2010)
  [1006.4186 [hep-ph]].

\bibitem{Endo:2011mc} 
  M.~Endo, K.~Hamaguchi, S.~Iwamoto and N.~Yokozaki,
  Phys.\ Rev.\ D {\bf 84}, 075017 (2011)
  [1108.3071 [hep-ph]].

\bibitem{Faroughy:2014oka}
  C.~Faroughy and K.~Grizzard,
  Phys.\ Rev.\ D {\bf 90} (2014) 3,  035024
  [arXiv:1405.4116 [hep-ph]].


\bibitem{Ellis:2014dza} 
  S.~A.~R.~Ellis, R.~M.~Godbole, S.~Gopalakrishna and J.~D.~Wells,
  JHEP {\bf 1409}, 130 (2014)
  [arXiv:1404.4398 [hep-ph]].

\bibitem{RaiseHiggs}
  J.~L.~Evans et al., 
  1108.3437 [hep-ph].
  T.~Moroi et al., 
  Phys.\ Lett.\ B {\bf 709}, 218 (2012)
  [1112.3142 [hep-ph]].
  M.~Endo et al., 
  Phys.\ Rev.\ D {\bf 85}, 095012 (2012)
  [1112.5653 [hep-ph]].
  S.~P.~Martin and J.~D.~Wells,
  Phys.\ Rev.\ D {\bf 86}, 035017 (2012)
  [1206.2956 [hep-ph]].
  M.~Endo et al., 
  JHEP {\bf 1301}, 181 (2013)
  [1212.3935 [hep-ph]].
  

\bibitem{Higgsmass} 
  D.~M.~Pierce, J.~A.~Bagger, K.~T.~Matchev and R.~-j.~Zhang,
  Nucl.\ Phys.\ B {\bf 491}, 3 (1997)
  [hep-ph/9606211].
  A.~Dedes and P.~Slavich,
  Nucl.\ Phys.\ B {\bf 657}, 333 (2003)
  [hep-ph/0212132].
  A.~Dedes et al., 
  Nucl.\ Phys.\ B {\bf 672}, 144 (2003)
  [hep-ph/0305127]. 
  A.~Brignole et al., 
  Nucl.\ Phys.\ B {\bf 643}, 79 (2002)
  [hep-ph/0206101]. 
  A.~Brignole et al., 
  Nucl.\ Phys.\ B {\bf 631}, 195 (2002)
  [hep-ph/0112177]. 
  G.~Degrassi et al., 
  Nucl.\ Phys.\ B {\bf 611}, 403 (2001)
  [hep-ph/0105096].


\bibitem{Staub:2013tta}
  F.~Staub,
  Comput.\ Phys.\ Commun.\  {\bf 185} (2014) 1773
  [arXiv:1309.7223 [hep-ph]]. \\
  F.~Staub,
  Comput.\ Phys.\ Commun.\  {\bf 182} (2011) 808
  [arXiv:1002.0840 [hep-ph]].

\bibitem{Martin:1993zk}
  S.~P.~Martin and M.~T.~Vaughn,
  Phys.\ Rev.\ D {\bf 50} (1994) 2282
   [Erratum-ibid.\ D {\bf 78} (2008) 039903]
  [hep-ph/9311340].
    
\bibitem{Ellis:2015}
  S.A.R. Ellis, J.D. Wells,
  arXiv:1502.01362 [hep-ph].
  
\bibitem{Peskin:1990zt}
  M.~E.~Peskin and T.~Takeuchi,
  Phys.\ Rev.\ Lett.\  {\bf 65} (1990) 964.\\
  M.~E.~Peskin and T.~Takeuchi,
  Phys.\ Rev.\ D {\bf 46} (1992) 381.

\bibitem{Martin:2004id}
  S.~P.~Martin, K.~Tobe and J.~D.~Wells,
  Phys.\ Rev.\ D {\bf 71} (2005) 073014
  [hep-ph/0412424].
  
\bibitem{Lavoura:1992np}
  L.~Lavoura and J.~P.~Silva,
  Phys.\ Rev.\ D {\bf 47} (1993) 2046.
  
\bibitem{Baak:2014ora}
  M.~Baak {\it et al.},
  arXiv:1407.3792 [hep-ph].
 
\bibitem{Baak:2013fwa}
  M.~Baak {\it et al.},
  arXiv:1310.6708 [hep-ph].

\bibitem{MSSM Higgs}
  A.~Djouadi,
  Phys.\ Rept.\  {\bf 459} (2008) 1
  [hep-ph/0503173].

\bibitem{HiggsHuntersGuide}
  J.~F.~Gunion, H.~E.~Haber, G.~L.~Kane and S.~Dawson,
  ``The Higgs Hunter's Guide,''
  Front.\ Phys.\  {\bf 80} (2000) 1.

\bibitem{CMS:2013xfa}
  [CMS Collaboration],
  arXiv:1307.7135.

\bibitem{Almeida:2013jfa}
  L.~G.~Almeida, S.~J.~Lee, S.~Pokorski and J.~D.~Wells,
  Phys.\ Rev.\ D {\bf 89} (2014) 3,  033006
  [arXiv:1311.6721 [hep-ph]].

\bibitem{Chatrchyan:2013uxa}
  S.~Chatrchyan {\it et al.}  [CMS Collaboration],
  Phys.\ Lett.\ B {\bf 729} (2014) 149
  [arXiv:1311.7667 [hep-ex]].

\bibitem{Bhattacharya:2013iea}
  S.~Bhattacharya, J.~George, U.~Heintz, A.~Kumar, M.~Narain and J.~Stupak,
  arXiv:1309.0026 [hep-ex].

\end{thebibliography}
\end{document}